\documentclass{article}
\usepackage{amssymb}
\usepackage{graphicx}
\usepackage{tabularx}
\usepackage{ltxtable}

\input{tcilatex}
\begin{document}

\title{Relativistic Quantum Games in Noninertial Frames}
\author{Salman Khan\thanks{%
sksafi@phys.qau.edu.pk}, M. Khalid Khan \\
Department of Physics, Quaid-i-Azam University, \\
Islamabad 45320, Pakistan.}
\maketitle

\begin{abstract}
We study the influence of Unruh effect on quantum non-zero sum games. In
particular, we investigate the quantum Prisoners' Dilemma both for entangled
and unentangled initial states and show that the acceleration of the
noninertial frames disturbs the symmetry of the game. It is shown that for
maximally entangled initial state, the classical strategy $\hat{C}$
(cooperation) becomes the dominant strategy. Our investigation shows that
any quantum strategy does no better for any player against the classical
strategies. The miracle move of Eisert et al \cite{Eisert} is no more a
superior move. We show that the dilemma like situation is resolved in favor
of one player or the other.\newline
\end{abstract}

\section{Introduction}

Quantum game theory, began from the seminal paper of Meyer \cite{Meyer}. It
deals with classical games in the domain of quantum mechanics. For the last
few years much valuable work has been done in this area. Various quantum
protocols have been developed and many classical games have been extended to
the domain of quantum mechanics. It has been shown that quantum
superposition and prior quantum entanglement between the players' states
ensure quantum players to outperform the classical counterparts through
quantum mechanical strategies[2-9]. Quantum entanglement is one of the
powerful tools of quantum mechanics and plays the role of a kernel in
quantum information and quantum computation. A prior quantum entanglement
between two spatially separated parties increases the number of classical
information communicated between them to twice the number of classical bits
communicated in the case of unentangled state \cite{Bennette,Brassard}.
Recently, the behavior of prior entanglement shared between two spatially
separated parties has been extended to the relativistic setup in noninertial
frames \cite{Alsing,Ling,Gingrich,Pan, Schuller, Terashima} and interesting
results have been obtained. Alsing \textit{et al.} \cite{Alsing} have shown
that the entanglement between the two modes of a free Dirac field is
degraded by the Unruh effect and asymptotically reaches a nonvanishing
minimum value in the limit of infinite acceleration.

In this paper, we study the influence of Unruh effect on the payoffs
function of the players in the quantum non-zero sum games. In particular, we
concentrate on the quantum Prisoners' Dilemma \cite{Eisert}. We show that
the payoffs function of the players are strongly influenced by the
acceleration of the noninertial frame and the symmetry of the game is
disturbed. It is shown that under some particular situations, the classical
strategy $\hat{C}$ becomes the dominant strategy and the classical strategy
profiles ($\hat{C},\hat{C}$) and ($\hat{D},\hat{D}$) are no more the Pareto
optimal and the Nash equilibrium, respectively. We show that the dominance
of the quantum player ceases in the presence of acceleration of the
noninertial frame. In the infinite limit of acceleration, new Nash
equilibrium arises. Furthermore, the dilemma like situation under every
condition, we consider here, is resolved in the favor of one player or the
other or both.

\section{The Prisoners' Dilemma}

The Prisoners' Dilemma is a well known non-zero sum game, which has a
widespread applications in many areas of science. Each one of the two
players (Alice and Bob) has to choose one of the two pure strategies
simultaneously. The two pure strategies are called cooperation ($C$) and
defection ($D$). The reward to the action of a player depends not only on
his own strategy but also on the strategy of his opponent. The classical
payoff matrix of the game has the structure given in Table $1$. The first
number in each pair of the matrix corresponds to Alice's payoff and the
second number in a pair to Bob's payoff. This is a symmetric noncooperative
game where each player tries to maximize his/her own payoff. The catch of
the dilemma is that $D$ is the dominant strategy, that is, rational
reasoning forces each player to defect, and thereby doing substantially
worse than if they would both decide to cooperate. The quantum form of the
Prisoners' Dilemma was studied for the first time by Eisert et al \cite%
{Eisert}.

\begin{table*}[htb]%
\caption{Payoff matrix for the classical Prisoners' Dilemma. The first entry in a pair
of numbers denotes the payoff of Alice and the second entry represents Bob's
payoff. \label{table:1}}$%
\begin{tabular}{|c|c|c|}
\hline\hline
& Bob: $C$ & Bob: $D$ \\ \hline\hline
Alice: $C$ & $3,3$ & $0,5$ \\ \hline\hline
Alice: $D$ & $5,0$ & $1,1$ \\ \hline\hline
\end{tabular}%
$

\end{table*}%

\section{Calculation}

We consider that Alice and Bob share an entangled initial state $|\psi
_{i}\rangle =\hat{J}|00\rangle _{A,B}$ of two qubits (one for each player)
at a point in flat Minkowski spacetime. The subscripts $A,B$ of the ket
stand, respectively, for Alice and Bob, which means that the first entry in
the ket corresponds to Alice and the second entry corresponds to Bob. The
unitary operator $\hat{J}$\ is an entangling operator and is given by%
\begin{equation}
\hat{J}=\mathrm{exp}[i\frac{\gamma }{2}\hat{D}_{1}\otimes \hat{D}_{1}],
\label{A}
\end{equation}%
where $\gamma \in \lbrack 0,\pi /2]$ and is a measure of the degree of
entanglement in the initial state. The initial state is maximally entangled
when $\gamma =\pi /2$. The operator $\hat{D}_{1}$ is given by%
\begin{equation}
\hat{D}_{1}=\left( 
\begin{array}{cc}
0 & 1 \\ 
-1 & 0%
\end{array}%
\right) ,
\end{equation}%
The entangling operator $\hat{J}$ must be symmetric with respect to the
interchange of the two players in order to execute a fair game and must be
known to both players for the knowledge of the degree of entanglement in the
initial state. The initial state, after the entangling operator is applied,
becomes%
\begin{equation}
|\psi _{i}\rangle =\cos \frac{\gamma }{2}|00\rangle _{A,B}+i\sin \frac{%
\gamma }{2}|11\rangle _{A,B}.  \label{1}
\end{equation}%
\begin{figure}[h]
\begin{center}
\begin{tabular}{ccc}
\vspace{-0.5cm} \includegraphics[scale=1.2]{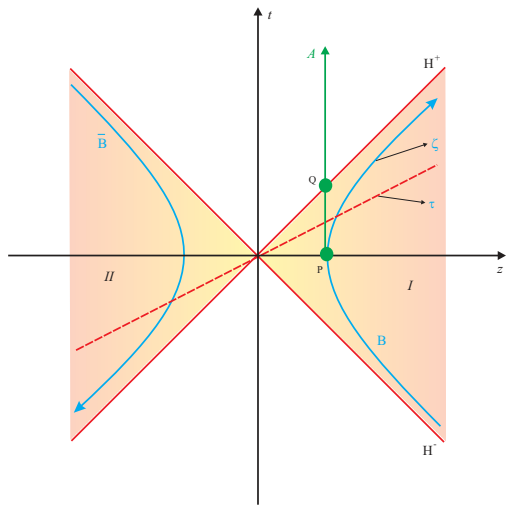}\put(-320,220) &  & 
\end{tabular}%
\end{center}
\caption{(color online) Rindler spacetime diagram: A uniformly accelerated
observer Bob ($B$) moves on a hyperbola with constant acceleration $a$ in
region $I$ and a fictitious observer anti-Bob ($\bar{B}$) moves on a
corresponding hyperbola in causally diconnected region $II$. The coordinates
\ $\protect\tau $ and $\protect\zeta $ are the Rindler coordinates in Bob's
frame, which represent constant proper time and constant position,
respectively. Lines $H^{\pm }$ are the horizons that represent Bob's future
and past and correspond to $\protect\tau =+\infty $ and $\protect\tau %
=-\infty $. Alice and Bob share an entangled initial state at point $P$ and $%
Q$ is the point where Alice crosses Bob's future horizon.}
\label{Figure1}
\end{figure}
We consider that Bob then moves with a uniform acceleration and Alice stays
stationary. Each player is equipped with a device which is sensitive only to
a single mode in their respective regions. To cover Minkowski space, two
different sets of Rindler coordinates ($\tau ,\xi $) (see Fig. ($1$)) that
differe from each other by an overall change in sign and define two Rindler
regions ($I,II$) are necessary (for detail see \cite{Alsing} and references
therein). A uniformly accelerated particle (observer) in one Rindler region
is causally disconnected from the other Rindler region at the opposite side.
Thus an observer in region $I$ has no access to the information that leaks
into region $II$. The opposite is true for an observer in region $II$. An
observer in region $II$ is called anti-observer (anti-particle) of the
observer in region $I$. The inaccessible information that leaks into the
opposite region is as the system is decohered. The decohrence effects in
quantum games in inertial frames are studied by a number of authors \cite%
{Chen,Flitney3,Salman}. Particularly, in Ref. \cite{Chen} the decoherence
effects on quantum Prisoners' Dilemma has been studied using various quantum
channels. However, the results of our calculations in the relativistic set
up of the game\ in noninertial frames are different from the one obtained in
Refs. \cite{Chen,Flitney3}. The creation operator ($a_{k}$) of particle and
annihilation operator ($b_{k}$) of antiparticle in Minskowski space are
related to the creation operator $c_{k}^{I}$ in region $I$ and annihilation
operator $d_{k}^{II\dag }$ in region $II$ by the following Bogoliubov
transformation%
\begin{equation}
\left( 
\begin{array}{c}
a_{k} \\ 
b_{k}^{\dag }%
\end{array}%
\right) =\left( 
\begin{array}{cc}
\cos r & -e^{-i\phi }\sin r \\ 
e^{i\phi }\sin r & \cos r%
\end{array}%
\right) \left( 
\begin{array}{c}
c_{k}^{I} \\ 
d_{k}^{II\dag }%
\end{array}%
\right) ,  \label{A1}
\end{equation}%
where $k$ represents a single mode in each region and $\phi $ is an
unimportant phase that can always be absorbed into the definition of the
operators and $r$ is the dimensionless acceleration parameter given by $\cos
r=\left( e^{-2\pi \omega c/a}+1\right) ^{-1/2}$. The constants $\omega $, $c$
and $a$, in the exponential stand, respectively, for Dirac particle's
frequency, speed of light in vacuum and Bob's acceleration. The parameter $%
r=0$ when acceleration $a=0$ and $r=\pi /4$ when $a=\infty $. We see that
the transformation in Eq. (\ref{A1}) mixes a particle in region $I$ and an
antiparticle in region $II$. A similar transformation exists for an
antiparticle's operator in region $I$ and a particle's operator in region $II
$ \cite{Alsing}. In fact, a given Minskowski mode of a particular frequency
spreads over all positive Rindler frequencies ($\omega /(a/c)$) that peaks
about the Minskowski frequency \cite{Takagi,Alsing2}. However, to simplify
our problem we consider a single mode only in the Rindler region $I$, an
approximation that results into Eq. (\ref{A1}). This is valid if the
observers' detectors are highly monochromatic that detects the frequency $%
\omega _{A}\sim \omega _{B}=\omega $.

From Eq. (\ref{A1}) one can find that%
\begin{equation}
a_{k}=\cos rc_{k}^{I}-e^{-i\phi }\sin rd_{k}^{II\dag }.  \label{A2}
\end{equation}%
From the accelerated Bob's frame, with the help of Eq. (\ref{A2}), one can
show that the Minkowski vacuum state is found to be a two-mode squeezed state%
\begin{equation}
|0\rangle _{M}=\cos r|0\rangle _{I}|0\rangle _{II}+\sin r|1\rangle
_{I}|1\rangle _{II}.  \label{2}
\end{equation}%
Note that in Eq. (\ref{2}) we put $I$ and $II$ in the subscript of the kets
to represent the Rindler modes in region $I$ and region $II$, respectively.
Eq. (\ref{2}) shows that the noninertial observer that moves with a constant
acceleration in region $I$ sees a thermal state instead of the vacuum state.
This effect is called the Unruh effect \cite{Davies,Unruh}. Similarly, using
the adjoint of Eq. (\ref{A2}) one can easily show that the excited state in
Minkowski spacetime is related to Rindler modes as follow%
\begin{equation}
|1\rangle _{M}=|1\rangle _{I}|0\rangle _{II}.  \label{3}
\end{equation}

In terms of Minkowski mode for Alice and Rindler modes for Bob, the
entangled initial state of Eq. (\ref{1}) by using Eqs. (\ref{2}) and (\ref{3}%
) becomes%
\begin{eqnarray}
|\psi \rangle _{A,I,II} &=&\cos \frac{\gamma }{2}\cos r|0\rangle
_{A}|0\rangle _{I}|0\rangle _{II}  \nonumber \\
&&+\cos \frac{\gamma }{2}\sin r|0\rangle _{A}|1\rangle _{I}|1\rangle
_{II}+i\sin \frac{\gamma }{2}|1\rangle _{A}|1\rangle _{I}|0\rangle _{II}.
\label{4}
\end{eqnarray}%
Since Bob is causally disconnected from region $II$, we must take trace over
all the modes in region $II$. This leaves the following mixed density matrix
between the two players,%
\begin{equation}
\rho _{A,BI}=\left( 
\begin{array}{cccc}
\cos ^{2}r\cos ^{2}\frac{\gamma }{2} & 0 & 0 & -i\cos r\cos \frac{\gamma }{2}%
\sin \frac{\gamma }{2} \\ 
0 & \cos ^{2}\frac{\gamma }{2}\sin ^{2}r & 0 & 0 \\ 
0 & 0 & 0 & 0 \\ 
i\cos r\cos \frac{\gamma }{2}\sin \frac{\gamma }{2} & 0 & 0 & \sin ^{2}\frac{%
\gamma }{2}%
\end{array}%
\right) .  \label{5}
\end{equation}

In the quantum Prisoners' Dilemma, the strategic moves of Alice and Bob are
unitary operators which are given by \cite{Eisert}%
\begin{equation}
\hat{U}_{N}(\alpha ,\theta )=\left( 
\begin{array}{cc}
e^{i\alpha _{N}}\cos \frac{\theta _{N}}{2} & i\sin \frac{\theta _{N}}{2} \\ 
i\sin \frac{\theta _{N}}{2} & e^{-i\alpha _{N}}\cos \frac{\theta _{N}}{2}%
\end{array}%
\right) ,  \label{B}
\end{equation}%
where, the subscript $N=A,B$ represent Alice and Bob, $\theta \in \lbrack
0,\pi ]$ and $\alpha \in \lbrack 0,2\pi ]$. If cooperation and defection are
associated with the state $|0\rangle $ and the state $|1\rangle $,
respectively, then the quantum strategy $\hat{C}$ corresponds to $\hat{U}%
_{N}(0,0)$ and the quantum strategy $\hat{D}$ corresponds to $\hat{U}%
_{N}(0,\pi )$. To ensure that the classical game be a subset of the quantum
one, Eisert et al. \cite{Eisert} argued that the operator $\hat{J}$ must
commute with the tensor product of any pair of the moves $\hat{C}$ and $\hat{%
D}$. Since fermionic system in noninertial frames is a physically realizable
system, we hope that the encoding of the game might be practically possible.
Once decisions are made, the final density matrix prior to the measurement
becomes \cite{Eisert}%
\begin{equation}
\rho =\hat{J}^{\dag }\left( \hat{U}_{A}\otimes \hat{U}_{B}\right) \rho
_{A,I}\left( \hat{U}_{A}^{\dag }\otimes \hat{U}_{B}^{\dag }\right) \hat{J},
\label{6}
\end{equation}%
where $\hat{J}^{\dag }$\ is applied to disentangle the final density matrix.
The expected payoffs of the players are then found by using the following
equation%
\begin{equation}
P_{N}^{j_{1}j_{2}}=\sum_{i}\$_{N}^{j_{1(i)}j_{2(i)}}\rho _{ii},  \label{7}
\end{equation}%
where $\rho _{ii}$ ($i\in \lbrack 0,1]$) are the diagonal elements of the
final density matrix and $\$_{N}^{j_{1}(i)j_{2}(i)}$ ($j_{1},j_{2}\in
\lbrack C,D]$) are the classical payoffs of the players from Table $1$.

\section{Results and discussion}

The payoffs of the players for unentangled initial state ($\gamma =0$), when
each of them is allowed to play one of the two classical strategies, that
is, $\hat{C}=\hat{U}_{N}(0,0)$ or $\hat{D}=\hat{U}_{N}(0,\pi )$, are given
in Table $2$. The payoffs become the function of $r$.

\begin{table*}[htb]%
\caption{The payoff matrix of the players' payoffs as a function of the
acceleration of Bob's frame. The first entry in every pair corresponds to
Alice's payoff and the second entry corresponds to Bob's payoff. The initial state of the game
is unentangled and the players are allowed to select a move from the two
pure classical moves. \label{table:1}}$%
\begin{tabular}{|c|c|c|}
\hline\hline
& Bob: $\hat{C}$ & Bob: $\hat{D}$ \\ \hline\hline
Alice: $\hat{C}$ & $3\cos ^{2}r,4-\cos 2r$ & $3\sin ^{2}r,4+\cos 2r$ \\ 
\hline\hline
Alice: $\hat{D}$ & $3+2\cos 2r,\sin ^{2}r$ & $3-2\cos 2r,\cos ^{2}r$ \\ 
\hline\hline
\end{tabular}%
$

\end{table*}%

One can easily see that the results of Table $2$ reduce to the classical
results of Table $1$ when the acceleration $a=0$ ($r=0$). The presence of
acceleration in the payoff functions of the players disturbs the symmetry of
the game. Neither the strategy profile ($\hat{C},\hat{C}$) nor the strategy
profile ($\hat{D},\hat{D}$) is an equilibrium outcome of the game in the
range of acceleration $0<r\leq \pi /4$. In this range of acceleration, Alice
always wins by playing $\hat{D}$ and always loses by playing $\hat{C}$. The
dilemma like situation is resolved in the favor of Alice. At infinite
acceleration ($r=\pi /4$), the strategy profiles ($\hat{C},\hat{C}$) $=$ ($%
\hat{C},\hat{D}$) $=$ ($3/2,4$), which means that if Alice plays $\hat{C}$,
Bob strategy becomes irrelevant and he wins all the time. Similarly, the
strategy profiles ($\hat{D},\hat{C}$) $=$ ($\hat{D},\hat{D}$) $=$ ($3,3/2$),
Alice is victorous, regardless of what strategy Bob executes. Non of the
strategy profiles is either Pareto optimal or Nash equilibrium.

However, for a maximal entangled state ($\gamma =\pi /2$), the situation is
entirely different. When both the players are restricted only to the
classical region of moves, the payoffs of the players for different strategy
profiles are given by%
\begin{eqnarray}
P_{A,B}^{CC} &=&1+\cos r+\cos ^{2}r+\frac{5}{4}\sin ^{2}r,  \nonumber \\
P_{A,B}^{DD} &=&\frac{1}{8}(17-8\cos r-\cos 2r),  \nonumber \\
P_{A}^{CD} &=&P_{B}^{DC}=\frac{1}{2}\cos ^{2}\frac{r}{2}(9+\cos r), 
\nonumber \\
P_{A}^{DC} &=&P_{B}^{CD}=\frac{1}{2}(9-\cos r)\sin ^{2}\frac{r}{2}.
\label{8}
\end{eqnarray}%
\begin{figure}[h]
\begin{center}
\begin{tabular}{ccc}
\vspace{-0.5cm} \includegraphics[scale=1.2]{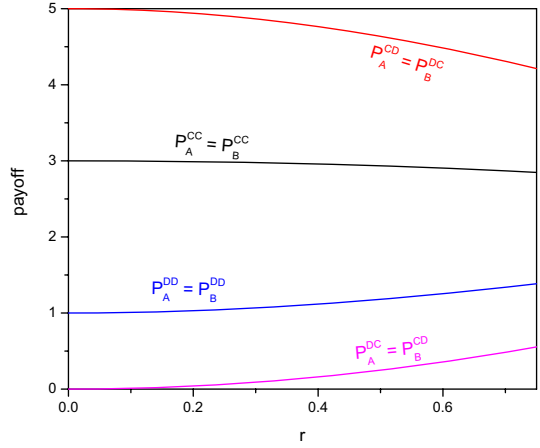}\put(-220,220) &  & 
\end{tabular}%
\end{center}
\caption{\textbf{(}color online) The payoffs for the maximally entangled
initial state are plotted against the acceleration parameter $r$ of Bob's
frame. The players are allowed to choose only the classical moves. The
subscripts stand for the players and the superscripts represent a strategy
profile.}
\label{Fig. 2}
\end{figure}
It can easily be seen from the payoffs function of Eq. (\ref{8}) that the
payoff matirx is symmetric and that for $r=0$, the classical results are
obtained. Also, the strategy profiles ($\hat{C},\hat{C}$) and ($\hat{D},\hat{%
D}$) are equilibrium points for the whole range of the acceleration of Bob's
frame. However, unlike the classical form and unentangled initial state of
the quantum form in inertial frames of the game, the strategy $\hat{C}$ in
this case becomes the dominant strategy and it always results in payoff $%
>2.83$ for all values of the acceleration of Bob's frame. Moreover, the
strategy profile ($\hat{C},\hat{C}$) becomes the Nash equilibrium and the
strategy profile ($\hat{D},\hat{D}$) becomes the Pareto optimal of the game
for all values of acceleration $a$. The payoffs of Eq. (\ref{8}), as
function of $r$ for all the possible strategy profiles, are plotted in Fig. $%
2$. It can be seen from the figure that playing $\hat{C}$ is the best option
for any player and hence resolves the dilemma like situation.

Now we consider the case in which the players are allowed to choose any
strategy from the allowed quantum mechanical strategic space. We first
consider the quantum strategy $\hat{Q}$ of Eisert \textit{et al.} \cite%
{Eisert}, which is given by%
\begin{equation}
\hat{Q}=\hat{U}\left( 0,\pi /2\right) =\left( 
\begin{array}{cc}
i & 0 \\ 
0 & -i%
\end{array}%
\right) .  \label{9}
\end{equation}%
The payoffs of the players when Alice chooses $\hat{Q}$ are given by%
\begin{equation}
P_{A,B}^{Q\theta _{B}}=\frac{1}{4}[9-\cos r((\cos r\mp 5)\cos \theta
_{B}+2\cos 2\alpha _{B}(\cos \theta _{B}+1)\pm 5)],  \label{11}
\end{equation}%
where $\theta _{B}=0$ or $\pi $ gives strategy $\hat{C}$ or strategy $\hat{D}
$ respectively. Now, if Bob plays $\hat{C}$, then $P_{A}^{QC}=P_{B}^{QC}$ is
an equilibrium point of the game. If Bob plays $\hat{D}$ then $%
P_{B}^{QD}=P_{A}^{CD}>P_{B}^{QC}>P_{A}^{QD}$ for all values of of the
acceleration of the Bob's frame. This means that the quantum strategy $\hat{Q%
}$ does no better for Alice against any of the two classical strategies of
Bob. In other words, $\hat{D}$ is the dominant strategy for Bob against
Alice strategy $\hat{Q}$. The same is true for Alice, if Bob plays the
quantum strategy $\hat{Q}$. In fact the strategy profile ($\hat{Q}$, $\hat{C}
$) or ($\hat{C}$, $\hat{Q}$) is a Pareto optimal outcome. However, if both
players execute $\hat{Q}$, the payoffs $P_{A}^{QQ}=P_{B}^{QQ}=P_{A,B}^{CC}$
and hence the strategy profile ($\hat{Q},\hat{Q}$) is the Nash equilibrium.

Finally we consider the unfair game and the effect of the \textit{miracle
move} of Eisert \textit{et} \textit{al.} \cite{Eisert}. That is, if one
player is restricted to the classical strategic space, then, in the case of
inertial frames, the quantum player outsmarts the classical player all the
time if he or she plays the miracle move $\hat{M}$,%
\begin{equation}
\hat{M}=\hat{U}\left( \frac{\pi }{2},\frac{\pi }{2}\right) =\frac{i}{\sqrt{2}%
}\left( 
\begin{array}{cc}
1 & 1 \\ 
1 & -1%
\end{array}%
\right) .  \label{12}
\end{equation}%
However, this is not true in the case of noninertial frames. Let Alice plays 
$\hat{M}$ and Bob is restricted to the classical strategies, the payoffs of
the players become%
\begin{eqnarray}
P_{A}^{M\theta _{B}} &=&\frac{1}{4}(-3\cos ^{2}r\sin \theta _{B}+\cos r(\sin
\theta _{B}-7)+9),  \nonumber \\
P_{B}^{M\theta _{B}} &=&\frac{1}{4}(7\cos ^{2}r\sin \theta _{B}+\cos r(\sin
\theta _{B}+3)+9).  \label{13}
\end{eqnarray}%
It can easily be checked that $P_{A}^{M\theta _{B}}<P_{B}^{M\theta _{B}}$
irrespective of what strategy Bob executes. This result is symmetric with
respect to the interchange of the players. That is, if Alice is restricted
to the classical strategies and Bob plays $\hat{M}$, then, the payoffs of
the players in Eq. (\ref{13}) interchage and $\theta _{B}$ is replaced with $%
\theta _{A}$. The quantum player should never go for playing the quantum
miracle move of the inertial frames. The dominance of quantum player over
the classical one ceases in the case of noninertial frames. However, the
miracle move $\hat{M}$ always results in a winning payoff against the
quantum move $\hat{Q}$. Logically, putting $r=0$ in Eq. (\ref{13}) should
give the results of quantum Prisoners' Dilemma in the inertial frames but
this is not so. Eq. (\ref{13}) gives inverted results when $r=0$, that is,
Alice's payoff becomes Bob's payoff of the inertial frame and vice versa. We
have no explanation for this inconsistency.

\section{Conclusion}

We study the influence of Unruh effect on the payoffs function of the
players in the quantum Prisoners' Dilemma. For unentangled initial state,
the Unruh effect gives rise to an asymmetric payoff matrix in contrast to
the payoff matrix for the classical form and quantum form in the inertial
frames of the game. It is shown that for unentangled initial state, Alice
wins all the time if she plays $\hat{D}$ and loses if she plays $\hat{C}$.
As a result non of the classical strategies profile is either Perato optimal
or Nash equilibrium. We have shown that the Unruh effect limits the
dominance of the quantum player. The classical moves $\hat{C}$ or $\hat{D}$
becomes dominant against the quantum moves depending on the initial state
entanglement. It is shown that the miracle move $\hat{M}$ of the inertial
frames becomes the worst move that always results in loss against any
classical move. Nevertheless, against the quantum move $\hat{Q}$, it always
gives a winning payoff. It is shown that the dilemma like situation is
resolved in favor of one or the other player or for both players depending
on the degree of entanglement in the initial state of the game.

\section{Acknowledgment}

Salman Khan is thankful to World Federation of Scientists, Geneva,
Switzerland, for partially supporting this work under the National
Scholarship Program for Pakistan\textbf{.}

{\Huge Figures Captions}\newline
Figure $1$\textbf{. }(color online) Rindler spacetime diagram: A uniformly
accelerated observer Bob ($B$) moves on a hyperbola with constant
acceleration $a$ in region $I$ and a fictitious observer anti-Bob ($\bar{B}$%
) moves on a corresponding hyperbola in causally diconnected region $II$.
The coordinates \ $\tau $ and $\zeta $ are the Rindler coordinates in Bob's
frame, which represent constant proper time and constant position,
respectively. Lines $H^{\pm }$ are the horizons that represent Bob's future
and past and correspond to $\tau =+\infty $ and $\tau =-\infty $. Alice and
Bob share an entangled initial state at point $P$ and $Q$ is the point where
Alice crosses Bob's future horizon.\newline
Figure $2$\textbf{.} \textbf{(}color online) The payoffs for the maximally
entangled initial state are plotted against the acceleration parameter $r$
of Bob's frame. The players are allowed to choose only the classical moves.
The subscripts stand for the players and the superscripts represent a
strategy profile.\newline

\end{document}